\documentclass[a4paper,fleqn,usenatbib]{mnras}

\usepackage[T1]{fontenc}
\usepackage{ae,aecompl}

\usepackage{graphicx}	
\usepackage{amsmath}	
\usepackage{amssymb}	

\title[A disc with a power-law stress-pressure]{Properties of an accretion disc with a power-law stress-pressure relationship}
\author[M. Shadmehri, F. Khajenabi, S. Dib \& S. Inutsuka]{
Mohsen Shadmehri$^{1,2}$,\thanks{E-mail: m.shadmehri@gu.ac.ir}
Fazeleh Khajenabi$^1$,
Sami Dib $^{3,4}$ 
Shu-ichiro Inutsuka$^{5}$
\\
$^{1}$Department of Physics, Faculty of Science, Golestan University, Gorgan 49138-15739, Iran\\
$^{2}$Research Institute for Astronomy and Astrophysics of Maragha (RIAAM), Maragha, P.O. Box: 55134-441, Iran\\
$^{3}$Max-Planck Institute for Astronomy, K\"{o}nigstuhl 17, D-69117, Heidelberg, Germany\\
$^{4}$Niels Bohr International Academy, Niels Bohr Institute, Blegdamsvej 17, DK-2100 Copenhagen, Denmark\\
$^{5}$Department of Physics, Nagoya University, Furo-cho, Chikusa-ku, Nagoya, Aichi, 464-8602, Japan \\
}

\date{Accepted XXX. Received YYY; in original form ZZZ}

\pubyear{2017}

\begin{document}
\label{firstpage}
\pagerange{\pageref{firstpage}--\pageref{lastpage}}
\maketitle

\begin{abstract}
Recent numerical simulations of magnetized accretion discs show that the radial-azimuthal component of the stress tensor due to the magnetorotational instability (MRI) is well represented by a power-law function of the gas pressure rather than a linear relation which has been used in most of the accretion disc studies. The exponent of this power-law function which depends on the net flux of the imposed magnetic field is reported in the range between zero and unity. However, the physical consequences of this power-law stress-pressure relation within the framework of the standard disc model have not been explored so far. In this study, the structure of an accretion disc with a power-law stress-pressure relation is studied using analytical solutions in the steady-state and time-dependent cases. The derived solutions are applicable to different accreting systems, and as an illustrative example, we explore structure of protoplanetary discs using these solutions. We show that the slopes of the radial surface density and temperature distributions become steeper with  decreasing  the stress exponent. However,  if the disc opacity is dominated by icy grains and value of the stress exponent is less than about $0.5$, the surface density and temperature profiles become so steep that make them unreliable. We also obtain analytical solutions for the protoplanetary discs which are irradiated by the host star. Using these solutions, we find that the effect of the irradiation becomes more significant with decreasing the stress exponent. 
\end{abstract}




\section{Introduction}
Theoretical attempts to understand the nature of the angular momentum transport and turbulence in accretion discs are still under active investigation and intense debate. Depending on the physical properties of an accreting system, various angular momentum transport mechanisms have been proposed over the recent decades \citep[e.g.,][]{Balbus91,Lovelace99,Stoll14,Rafikov15}. In the weakly ionized accretion discs,  magnetorotational instability \citep[MRI;][]{Balbus91} is believed to be very efficient in transporting angular momentum. However, in the regions of a disc where the level of ionization is very low, such as the outer parts of a protoplanetary disc (PPD), gravitational instability has been proposed as the dominant mechanism for the angular momentum transport  \citep[e.g.,][]{Rice2005,Cossins,Rafikov09}.

Owing to the non-linear nature of the turbulence,  numerical simulations play a vital role in our understanding of the angular momentum transport in accretion discs. Analytical models, however,  are also very useful for describing the structure of the accretion discs due to the simplicity in interpreting the results and the possibility of examining a wider range of the input parameters. In these simplified models, turbulence is described in terms of an effective viscosity, which is prescribed in an {\it ad hoc} fashion.  In the standard theory of accretion discs  \citep{Shakura,Lynden}, the radial-azimuthal component of the stress tensor, $\Pi_{ r\phi}$, is assumed to be proportional to the gas pressure and the coefficient of the proportionality, $\alpha$, is a dimensionless parameter with a value less than unity. This form of the stress tensor and the resulting viscosity that is known as $\alpha -$formalism \citep{Shakura} is defined independently  of the angular momentum transport mechanism. Many authors have then tried to justify the $\alpha$-formalism using analytical methods or numerical simulations \citep[e.g.,][]{Balbus99}. 

In some of the accreting systems, like a disc around a compact object, not only the gas pressure, but also the radiation pressure are dynamically important. However, it is not clear if the stress tensor is still proportional to the gas pressure or total pressure. For this reason and in the light of some theoretical arguments, some authors investigated the properties of the discs with a stress tensor proportional to a power-law function of a combination of the gas and total pressures \citep[e.g.,][]{Taam,Szu,Merloni}. Another line of research is related to the thermal stability of these disc models which leads to some restrictions on the proposed viscosity \citep[e.g.,][]{Lightman,Shakura76}. 

The stress parameter $\alpha$ is commonly assumed to be a fixed input parameter, however, there are theoretical arguments that $\alpha$ can be a function of the spatial coordinates or even disc quantities depending on the angular momentum transport mechanism. For instance, when the MRI is efficient, it has been argued that $\alpha$ is a power-law function of the magnetic Prandtl number \citep{Fromang,Lesur,Simon}.  \cite{Takahashi} constructed steady-state thin disc models by allowing a power-law dependence  of the stress parameter $\alpha$ on the magnetic Prandtl number. It has also been suggested that in a disc with MRI-driven turbulence, the toroidal component of the magnetic field is so strongly amplified that the magnetic pressure exceeds the total pressure in the disc. Properties of such a magnetically dominated accretion disc have been studied by \cite{Begelman} and their solutions are stable subject to the thermal and viscous instabilities.  In the discs with the gravity-driven turbulence, on the other hand, the stress parameter is shown to be a complicated function of the disc quantities \cite[e.g.,][]{Gammie2001,Goodman2003,Rice2005,Rafikov09,Cossins,shadmehri17}. In self-gravitating discs, however, other non-standard viscosity prescriptions have been proposed \citep[e.g.,][]{Duschl}. 

Thus, a key issue in constructing analytical models for the structure of accretion discs is the true dependence of $ \Pi_{ r\phi}$ on the disc quantities. Most of the previous numerical studies tried to determine the value of the stress parameter $\alpha$. It is not clear, however, if the radial-azimuthal component of the stress tensor is linearly proportional to the gas pressure \citep[e.g.,][]{Sano,Minoshima}. Recently, \cite{Ross} performed an interesting study using local numerical simulations with unstratified boxes to test whether $\Pi_{ r\phi}$ is proportional to the gas pressure. Although a few previous studies found a weak stress-pressure relationship, \cite{Ross} showed that there is a power-law relationship as $\Pi_{ r\phi} \propto p^n$, where $p$ is the gas pressure and the exponent $n$ is between 0 and 1. They argued that previous numerical simulations were box-size limited and suffered from an insufficiently large spatial range. The turbulent eddies, thereby, were restricted by the numerical domain instead of disc scale height. \cite{Ross} performed numerical simulations with a box-size larger than the disc scale-height and showed that the exponent $n$ is determined by the imposed magnetic flux and possible physical diffusivities.  

The purpose of this work is to construct steady-state and time-dependent models for the structure of an accretion disc with a power-law stress-pressure relationship and to explore its consequences using our analytical solutions. In the next section, we construct a steady-state disc model using a power-law stress-pressure relation. We find that the radial profile of the disc quantities strongly depends on the adopted stress exponent $n$. We also show that when this exponent is less than around 0.5, the derived solutions become unrealistic. In section 3, we generalize the solutions by considering radiative heating. The time-dependent analytical solutions  are obtained in section 4. We then investigate the behavior of the disc quantities using time-dependent solutions and constrain the stress exponent. We conclude with a summary of the results and directions for possible future works in  section 5.

\section{Steady-State Model}
\subsection{Basic Equations}
The basic equation for the time evolution of a thin accretion disc with a Keplerian rotation profile is \citep[e.g.,][]{frank}
\begin{equation}
\frac{\partial\Sigma}{\partial t}=\frac{3}{r}\frac{\partial}{\partial r} \left [ r^{1/2} \frac{\partial}{\partial r} \left (\nu\Sigma r^{1/2} \right ) \right ],
\end{equation}
where $\Sigma$ is the surface density and $\nu$ is the kinematic viscosity. In the steady-state case, and subject to the zero torque at the disc inner edge \citep{frank}, the above equation for the regions far from the inner boundary reduces to
\begin{equation}\label{eq:m1}
\dot{M}=3\pi\nu \Sigma ,
\end{equation}
where $\dot{M}$ is the accretion rate.  

Our goal is to examine the properties of a steady-state disc model with a stress tensor proportional to a power-law function of the gas  pressure, i.e. $\Pi_{ r \varphi} \propto p^n$, where $0<n\leq 1$. Thus, we can write
\begin{equation}\label{eq:stress}
\Pi_{ r\varphi} = -\alpha p_c (\frac{p}{p_c})^n ,
\end{equation}
where $p_c$ is a reference pressure and as we show later its value affects the disc structure. 
Using a relation between kinematic viscosity $\nu$ and the radial-azimuthal component of the stress tensor, i.e. $H \Pi_{ r\varphi}=\nu \Sigma r (d\Omega /dr) $,  we then obtain
\begin{equation}\label{eq:visco}
\nu = \frac{2}{3} \alpha p_c (\frac{p}{p_c})^n \frac{c_{\rm s}}{\Sigma \Omega^2},
\end{equation}
where $\Omega = \sqrt{GM_{\star} /r^3}$ is  the Keplerian angular velocity, and, $M_{\star}$ denotes the mass of the central star. Here, $r$ is the radial distance and $c_{\rm s}$ is  the sound speed. The disc thickness is $H=c_{\rm s}/\Omega$. 

The second main equation is obtained from the energy balance. Following many previous studies, we assume that the radiative cooling is balanced with the generated heat due to the turbulence only along the vertical direction:
\begin{equation}\label{eq:m2}
\sigma T_{\rm e}^4 = \frac{9}{8} \frac{\dot{M}}{3\pi} \Omega^2 ,
\end{equation} 
where $\sigma$ is Stephan-Boltzmann constant and $T_{\rm e}$ is the effective surface temperature. The right-hand side term represents viscous heating. However, there are more heating sources which can be implemented here. For instance, in protoplanetary discs (PPDs), the heating due to cosmic rays and irradiation of the central star play a vital role. Although these processes can be considered in a more sophisticated model, we consider radiative heating using a simplified model in  section 3. Furthermore, one should note that {\it realistic} hydrodynamics simulations show that the radiative flux in the radial direction can be important in the disc \citep{Tsuka}. In order to proceed analytically, we construct a disc model following the standard approach. 

If the midplane temperature is denoted by $T$, the equation of radiative transfer in the optically thick regime is simplified to the following equation:
\begin{equation}\label{eq:m3}
T^4 = \frac{3}{16} \tau  T_{\rm e}^4,
\end{equation}
where $\tau$ is the optical depth, i.e. $ \tau =\kappa \Sigma $ and $\kappa$ is the opacity.  The opacity $\kappa$ is a complicated function of the density and temperature. We approximate it as a power-law function of the temperature \citep[e.g.,][]{Bell94}:
\begin{equation}\label{eq:opacity}
\kappa = \kappa_0 T^{\beta},
\end{equation}
where the exponent $\beta$ depends on the adopted temperature interval. For instance, at low temperature $T\lesssim 170$, the opacity due to ice grains is characterized by \citep[][]{Bell94} 
\begin{equation}
\beta =2, \hspace{0.5cm} \kappa_{0}= 5\times 10^{-4} {\rm cm}^2 {\rm g}^{-1} {\rm K}^{-2}.
\end{equation}
At higher temperatures, ice evaporation  occurs and the opacity is approximated by $\kappa \simeq 0.1 T^{1/2}$ cm$^{2}$ g$^{-1}$ \citep[][]{Bell94}. Our focus is to explore the disc properties with $\beta=2$, however, we derive general disc solutions and the discussions can be trivially extended to other opacity regimes. Our main findings of the stress exponent dependence of the solutions are independent of the adopted opacity regime. However, the quantitative trends of the solutions depend on the opacity exponent $\beta$. 

In the inner region of a PPD where the temperature is greater than $10^3$ K, thermal ionization is enough to activate the MRI. These regions with a very high surface density are optically thick in the vertical direction. In the outer parts of a PPD, nevertheless, MRI is activated with cosmic-ray ionization. It then requires that the surface density should be smaller than $10^2$ g cm$^{-2}$, and, regions with the surface density smaller than this value are optically thin. Therefore, we restrict our analysis to the optically thick regime. 

\subsection{Optically thick solutions}
The above algebraic equations can be solved analytically. To do so, we first rewrite equation (\ref{eq:visco}) for the viscosity as follows
\begin{equation}
\nu = \alpha \nu_0 \left (\frac{\Sigma}{\Sigma_0} \right )^{n-1} \left (\frac{c_{\rm s}}{c_{\rm s0}} \right )^{n+1} \left ( \frac{r}{r_0}\right )^{3(2-n)/2},
\end{equation}
where  $\Sigma_0$, $c_{\rm s0}$ and $r_0$ are the reference surface density, sound speed and radial distance, respectively. Furthermore, we have $\nu_0 = (2/3)c_{\rm s0}^{2} \Omega_{0}^{-1} \chi^{n-1} (M_{\star}/M_{\odot})^{(n-2)/2}$ and we have $\Omega_0 =\sqrt{GM_{\odot}/r_{0}^3}$. Here, we introduced $\chi=p_0 /p_c$ and it is treated as a model parameter. 

Thus, equation (\ref{eq:m1}) becomes 
\begin{equation}\label{eq:m1-b}
\frac{\dot{M}}{\dot{M}_0}=\alpha \xi_1 \left (\frac{M_\star}{M_\odot} \right )^{(n-2)/2} \left (\frac{\Sigma}{\Sigma_0} \right )^n \left (\frac{c_{\rm s}}{c_{\rm s0}} \right )^{n+1} \left ( \frac{r}{r_0}\right )^{3(2-n)/2},
\end{equation}
where $\dot{M}_{0}$ is a reference accretion rate and $\xi_1$ is a dimensionless parameter which depends on the reference quantities: $\xi_1 = 3\pi \nu_0 \Sigma_0 / \dot{M}_0 $. Using equation (\ref{eq:m3}), furthermore, we can rewrite equation (\ref{eq:m2}) as follows
\begin{equation}\label{eq:m2-b}
\left (\frac{T}{T_0} \right )^{4-\beta} \left (\frac{\Sigma}{\Sigma_0} \right )^{-1} = \xi_2 \xi_3 \left (\frac{\dot{M}}{\dot{M}_0} \right ) \left (\frac{M_\star}{M_\odot} \right ) \left (\frac{r}{r_0} \right )^{-3},
\end{equation} 
where $T_0$ is a reference temperature and the  dimensionless parameters $\xi_2$ and $\xi_3$ are defined as $\xi_2 = (3/16)\kappa_0 T_{0}^\beta \Sigma_0 $ and $\xi_3 = 3\dot{M}_0 \Omega_{0}^2 / 8\pi \sigma T_{0}^4  $. On the other hand, the sound speed is $c_{\rm s}=(k_{\rm B} T/\mu m_{\rm H})^{1/2}$, where $k_{\rm B}$, $\mu$, and, $m_{\rm H}$ are the Boltzmann constant, mean molecular weight and mass of the Hydrogen, respectively. Therefore, 
\begin{equation}\label{eq:cs}
c_{\rm s} = c_{\rm s0} \left (\frac{T}{T_0} \right )^{1/2},
\end{equation}
where we assumed that $c_{\rm s0}=(k_{\rm B} T_0 / \mu m_{\rm H})^{1/2}$. Upon substituting the above equation into equation (\ref{eq:m1-b}) and using equation (\ref{eq:m2-b}), the surface density and temperature are obtained in terms of the radius and the disc parameters:
\begin{equation}\label{eq:dens-a}
\frac{\Sigma}{\Sigma_0} = {\cal A} \alpha^{\delta_1} \left (\frac{\dot{M}}{\dot{M}_0} \right )^{\delta_2} \left ( \frac{M_\star}{M_\odot}\right )^{\delta_3} \left (\frac{r}{r_0} \right )^{\delta_4},
\end{equation}
\begin{equation}\label{eq:temp-a}
\frac{T}{T_0}= {\cal B} \alpha^{\delta_5} \left (\frac{\dot{M}}{\dot{M}_0} \right )^{\delta_6 }  \left ( \frac{M_\star}{M_\odot}\right )^{\delta_7} \left (\frac{r}{r_0} \right )^{\delta_8}, 
\end{equation}
where the dimensionless parameters ${\cal A}$ and ${\cal B}$ are
\begin{displaymath}
{\cal A} = \xi_{1}^{-2(4-\beta)/[(9-2\beta )n+1]} (\xi_2 \xi_3 )^{-(n+1)/[(9-2\beta )n+1]}, 
\end{displaymath}
\begin{equation}
{\cal B} =\xi_{1}^{-2/[1+(9-2\beta)n]} (\xi_2 \xi_3 )^{2n/[1+(9-2\beta)n]},
\end{equation}
and the exponents $\delta_i$ $(i=1...8)$ are defined as
\begin{displaymath}
\delta_1 = -\frac{2(4-\beta )}{(9-2\beta )n+1}, \delta_2 = \frac{-n+7-2\beta}{(9-2\beta)n+1}, 
\end{displaymath}
\begin{displaymath}
\delta_3 = \frac{(\beta -5)n +7 -2\beta}{(9-2\beta)n+1}, \delta_4 = -3\delta_3 .
\end{displaymath}
\begin{displaymath}
\delta_5 = -\frac{2}{1+(9-2\beta)n}, \delta_6 =\frac{2(1+n)}{1+(9-2\beta)n}
\end{displaymath}
\begin{equation}
\delta_7 = \frac{2+n}{1+(9-2\beta )n}, \delta_8 = -\frac{6+3n}{1+(9-2\beta )n}.
\end{equation}

If we set $\beta=-0.01$ and $n=1$, the above solutions reduce to the results of \cite{Martin13}. The exponent $\beta$ is adopted depending on the considered temperature interval \citep{Ruden1991,Bell94}. If the opacity is assumed to be independent of temperature (i.e., $\beta = 0$), we then have $\kappa_0 = 3$ cm$^2$ g$^{-1}$ \citep{Bell94,Martin13}. Here, as we mentioned before, our focus is to explore solutions with opacity due to the icy dust particles $(\beta =2 )$ which is a reasonable approximation for the regions with a low temperature  $(T\lesssim 170 \hspace{1mm}{\rm K})$. For higher temperatures, i.e. $170 \hspace{1mm} {\rm K} \lesssim T \lesssim 1380 \hspace{1mm} {\rm K}$, the silicate and iron grains dominate and the opacity exponent becomes $\beta = 3/4$. The silicate and iron grains sublimate opacity regime correspond to the temperatures higher than about $1380$ K with the opacity exponent $\beta =-14$ \cite[][]{Ruden1991,Stepinski,Eksi}.  In our study, different values of $n$ are considered to explore how the dependence of the stress tensor on the gas pressure would affect the  behavior of the disc quantities. 

\subsubsection{Behavior of the optically thick Steady-State Solutions}

\begin{figure}
\includegraphics[scale=0.6]{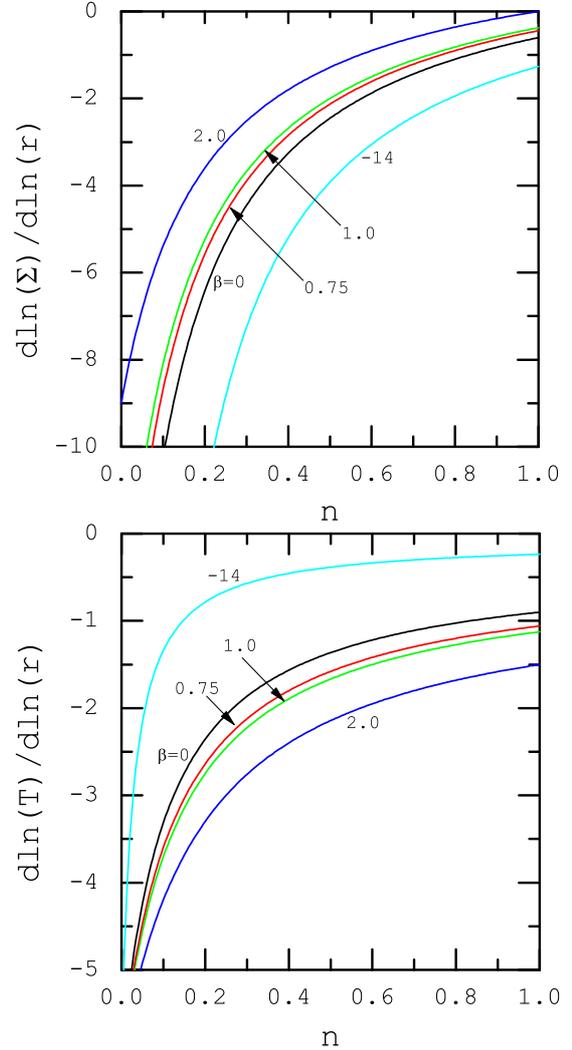}
\caption{The radial slope of the surface density (top) and the temperature (bottom) in the optically thick regime as a function of the exponent $n$. Each curve is labelled with the corresponding opacity exponent $\beta$.}\label{fig:f1}
\end{figure}

To assess the dependence of the surface density and temperature on the stress exponent $n$, Figure \ref{fig:f1} displays the radial slope of the disc quantities as a function of the exponent $n$ for different values of the parameter $\beta$. An obvious feature of this figure is that the radial slopes of both the surface density and temperature strongly depend on the exponents $n$ and $\beta$. This helps us to constrain the exponent $n$ in light of the adopted density or temperature profiles which have been implemented by the current studies on the  PPDs. For instance, a commonly used model is the minimum mass solar nebula \citep[MMSN;][]{hayashi81}, in which the surface density and temperature are approximated as $\Sigma \propto r^{-3/2}$ and $T\propto r^{-1/2}$. Our temperature profile is generally steeper than the temperature distribution of the MMSN, however, if  the density distribution is used as a diagnostic, there is a certain value of the stress exponent $n$ for which the density slope becomes $-3/2$. If we set $\delta_4 = -3/2$, then the corresponding value of $n$ for a given exponent $\beta$ becomes $n=(\beta -3.25)/(\beta - 4.75)$. Therefore, if we set $\beta=0$, 0.75, 1, 2, and -14 then acceptable values of $n$ become 0.68, 0.62, 0.60, 0.45, and, 0.92 respectively. 

We also note that the surface density of the PPDs can not be measured directly because their mass is dominated by H$_2$ which does not readily emit. The dust component of the PPDs, however, is the most commonly used tracer for determining their surface densities. Just recently, nevertheless, \cite{Powell} proposed  a novel method for measuring surface density of the PPDs by utilizing dust dynamics. \cite{Andrews2010} presented a detailed survey of 17 PPDs in the Ophiuchus star-forming region and found that the surface density has an exponent $-1$ that is flatter than the MMSN model. \cite{Barenfeld2017} investigated the properties of 57 circumstellar disks in the Upper Scorpius OB Association observed with ALMA at submillimeter wavelengths. They also found that the radial profiles in their sample are consistent with a slope -1. Interestingly, discs formed in the simulations of star cluster formation exhibit a similar radial surface density scaling \citep{Bate2018}. We, therefore, consider surface density profiles with a slope steeper than about -1.5 as unacceptable solutions. Figure \ref{fig:f1} shows that for an   opacity exponent $\beta=2$ (ice grains opacity regime) and $n\lesssim 0.5$, the surface density slope becomes much steeper than about -1.5 which makes these solutions unrealistic. For $\beta=3/4$ (silicate and iron grains opacity regime) and $n\lesssim 0.62$, the surface density slope becomes steeper than -1.5, but for $\beta=-14$ (silicate and iron grains sublimate opacity regime), the surface density slope becomes unrealistic for $n\lesssim 0.92$. 

\begin{figure}
\includegraphics[scale=0.6]{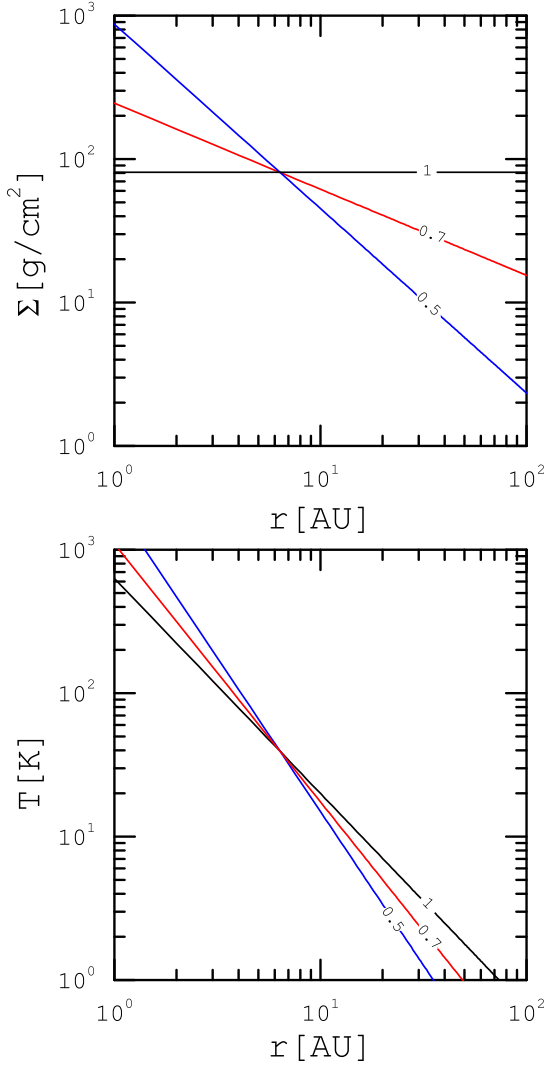}
\caption{Profiles of the optically thick solutions as a function of the radial distance. Top plot shows surface density and bottom plot displays temperature. Each curve is labeled with the corresponding exponent $n$. Here, we have $M_{\star}=M_{\odot}$, $\dot{M}=10^{-8} {\rm M}_{\odot}/{\rm yr}$, $\alpha=0.01$, $\Sigma_0 = 10$ g cm$^{-2}$ and $T_0 =10$ K, and, the opacity exponent is $\beta=2$.}\label{fig:f2}
\end{figure}

\begin{figure}
\includegraphics[scale=0.6]{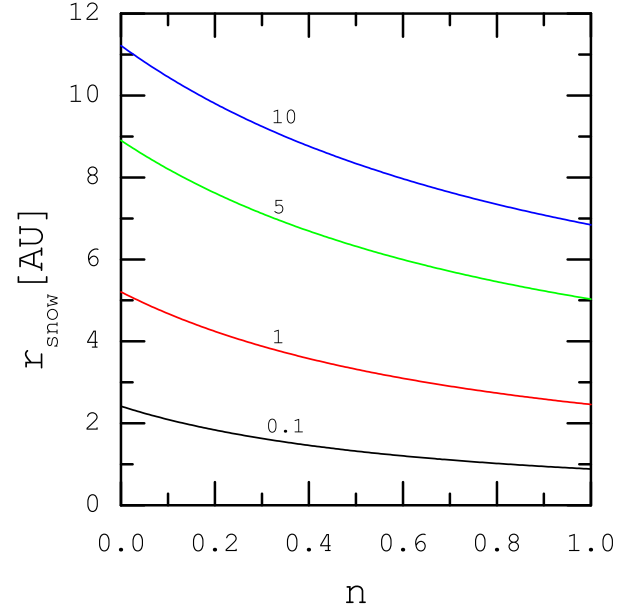}
\caption{Location of the snow-line as a function of the exponent $n$ is shown using optically thick solutions with $\beta=2$. Each curve is labeled with a  dimensionless accretion rate, i.e. $\dot{M}/(10^{-8} {\rm M}_{\odot} {\rm yr}^{-1})$. The viscosity coeffcient is $\alpha=0.01$, and, the central mass is $M_{\star}=M_\odot$.}\label{fig:f3}
\end{figure}

Now, we can adopt certain values for the reference quantities: $M_{\star}=M_{\odot}$, $\Sigma_0 = 10$ g ${\rm cm}^{-2}$, $T_0 = 10$ K, $r_0 =1$ AU, $\dot{M}_0 =10^{-8}$ M$_{\odot}$/yr, and, $\mu=2.1$. Given these values, and for $\beta =2$, we obtain $\xi_1 = 0.196 \chi^{n-1}$, $\xi_2 = 0.09$, and, $\xi_3 =5263$. From equations (\ref{eq:dens-a}) and (\ref{eq:temp-a}) and for $\chi=1$, we then obtain
\begin{displaymath}
\Sigma = 10 \exp [f_1 (n)] (\frac{\alpha}{0.01})^{-4/(1+5n)} (\frac{\dot{M}}{10^{-8} {\rm M}_{\odot}/{\rm yr}})^{(3-n)/(1+5n)} 
\end{displaymath}
\begin{equation}\label{eq:dens-b}
(\frac{M_\star}{M_\odot})^{3(1-n)/(1+5n)} (\frac{r}{\rm AU})^{-9(1-n)/(1+5n)} \hspace{0.2cm} {\rm g} \hspace{0.1cm} {\rm cm}^{-2},
\end{equation}
and
\begin{displaymath}
T = 10 \exp [f_2 (n) ] (\frac{\alpha}{0.01})^{-2/(1+5n)} (\frac{\dot{M}}{10^{-8} {\rm M}_{\odot}/{\rm yr}})^{2(1+n)/(1+5n)} 
\end{displaymath}
\begin{equation}\label{eq:temp-b}
(\frac{M_\star}{M_\odot})^{(2+n)/(1+5n)} (\frac{r}{\rm AU})^{-3(2+n)/(1+5n)} \hspace{0.2cm} {\rm K},
\end{equation}
where functions $f_1 (n)$ and $f_2 (n)$ are written as
\begin{equation}
f_1 (n) = \frac{18.7+6.2n}{1+5n},
\end{equation}
\begin{equation}
f_2 (n) = \frac{12.4(1+n)}{1+5n}.
\end{equation}

Using equations (\ref{eq:dens-b}) and (\ref{eq:temp-b}), Fig. \ref{fig:f2} depicts the effect of changing the exponent $n$ on the radial profiles of the surface density and temperature. The central part of a disc becomes denser and hotter as the stress exponent $n$ reduces. At a given radius, the enhancement of the surface density due to lowering $n$, however, is more pronounced compared to the enhancement of the temperature. Although  the stress exponent dependence of the surface density is stronger than the temperature distribution, we can see a significant variation of the temperature profile as the exponent $n$ varies. Thus, one can expect modification to the location of the snow-line. 

\subsubsection{Location of the Snow-Line}
The snow-line is defined as a radius where the  temperature becomes around $T_{\rm snow}=170$ K.  With equation (\ref{eq:temp-a}) for the temperature distribution in hand, the location of the snow-line becomes
\begin{equation}\label{eq:snow}
x_{\rm snow}=\tilde{T}_{\rm snow}^{1/\delta_8} {\cal B}_{\rm thick}^{-1/\delta_8} \alpha^{-\delta_5 / \delta_8 } \dot{m}^{-\delta_6 / \delta_8} M_{1}^{-\delta_7 /\delta_8},
\end{equation}
where $\tilde{T}_{\rm snow}=T_{\rm snow}/T_0 = 17$. Assuming $\beta =2$, this equation can be re-written as follows
\begin{displaymath}
r_{\rm snow}=\exp [f_3 (n)] (\frac{\alpha}{0.01})^{-2/3(2+n)} 
\end{displaymath}
\begin{equation}\label{eq:snow-b}
\times (\frac{\dot{M}}{10^{-8} {\rm M}_{\odot}/{\rm yr}})^{2(1+n)/3(2+n)}   (\frac{M_\star}{M_{\odot}})^{1/3} \hspace{0.2cm} {\rm AU},
\end{equation}
where
\begin{displaymath}
f_3 (n)=\frac{3(0.2+n)(5.5-n)}{(1+5n)(2+n)}.
\end{displaymath}

Figure \ref{fig:f3} displays the location of the snow-line as a function of the exponent $n$ for different values of a dimensionless accretion rate, i.e. $\dot{M}/(10^{-8} {\rm M}_{\odot} {\rm yr}^{-1})$. Here, we adopt $\alpha=0.01$ and $M_{\star}=M_\odot$. Obviously, a higher accretion rate implies a larger radius for the location of the snow-line. As the stress exponent $n$ decreases, the location of the snow-line shifts towards larger radii. However, the migration of $r_{\rm snow}$ is more significant for the higher accretion rates. However, the dependence of $r_{\rm snow}$ on the central mass, is a power-law relation with the exponent $1/3$. This result is independent of the stress exponent $n$. 

If we set $n=1$ and $\beta=-0.01$, our solution for the location of the snow-line reduces to the result of \cite{Martin13} who investigated  evolution of the snow-line in a PPD with (and without) dead zone and showed that the location of the snow-line is at a much larger distance from the central star when the dead zone is included. A fully turbulent model, however, predicts that the snow-line is actually even closer to the star. But all these predictions relied on a linear stress-pressure relation. When the  accretion rate is $10^{-8} {\rm M}_{\odot} {\rm yr}^{-1}$, Fig. \ref{fig:f3} shows that the snow-line migrates to radii larger than the Earth's orbit when the stress exponent is less than unity.

\subsection{Why all curves cross at a certain location?}

All solutions presented here exhibit a very evident feature: For a given set of input parameters, all curves cross at a certain radial location independent of $n$. This general trend can be understood in terms of our prescribed stress-pressure equation (\ref{eq:stress}) with two parameters $n$ and $p_c$. In all previous figures, for simplicity, we assumed that $\chi=1$ with fixed values for $\dot{M}$, $\alpha$ and $p_{c}$. Obviously, the disc pressure varies with the distance and it becomes equal to the given value $p_c$ at a certain radial distance. The lines will cross wherever the pressure in the disc is equal to $p_c$. At this location, the viscosity in each disc model is equal no matter what the value of $n$ is and so the surface density and temperature must be the same and hence the lines all cross at one location. We can, however, effectively choose the location of the lines crossing by choosing $p_c$. If we adopt a larger value for $\chi$, it means that the reference pressure $p_c=p_0 /\chi$ is smaller and thereby the curves intersect at a larger radius. We actually confirmed this trend using our analytical solutions. However, the general behavior of the solutions is similar to what we have presented in the figures for $\chi=1$.

\subsection{Stability of the solutions}
We now turn our attention to investigate whether our solutions are unstable subject to thermal instability or magnetic Prandtl number instability. Although our formalism can be generalized to include radiation pressure, we only considered gas dominated discs which are relevant for the PPDs. \cite{Takahashi} and \cite{Potter} derived a thermal-viscous instability criterion for a standard thin disc with a viscosity parameter that is a function of the magnetic Prandtl number or equivalently, of density and temperature. Here, we derive the thermal-viscous instability criterion for a standard thin disc with a power-law stress-pressure relationship. The instability criterion can be written as \citep{frank}
\begin{equation}\label{eq:cond}
\frac{\partial T_e}{\partial\Sigma}<0.
\end{equation}
Using the energy balance equation and power-law stress-pressure relation and at a given radius, we obtain
\begin{equation}\label{eq:ins-1}
T_{e}^4 \propto \alpha T^{\frac{n+1}{2}} \Sigma^n,
\end{equation}
or
\begin{equation}\label{eq:ins-2}
4T_{e}^3 \frac{\partial T_e}{\partial\Sigma} \propto \frac{\partial}{\partial\Sigma} \left (\alpha T^{\frac{n+1}{2}} \Sigma^n \right ).
\end{equation}
Upon substituting equation (\ref{eq:m3}) into equation (\ref{eq:ins-1}), we obtain
\begin{equation}\label{eq:ins-3}
T \propto \tau^{\frac{2}{7-n}} \Sigma^{\frac{2n}{7-n}}.
\end{equation} 
Using equations (\ref{eq:ins-1}), (\ref{eq:ins-2}) and (\ref{eq:ins-3}), the instability  criterion (\ref{eq:cond}) is simplified to the following inequality:
\begin{equation}
\frac{n+1}{8n} \frac{\partial \ln \tau}{\partial \ln \Sigma} + 1 <0.
\end{equation}
It is worth noting that for $n=1$, the above criterion reduces to the inequality (27) in \cite{Potter}. Using equation (\ref{eq:opacity}), a power-law relation between $\tau$ and $\Sigma$ is obtained, i.e.
\begin{equation}
\frac{\partial \ln \tau}{\partial \ln \Sigma}=\frac{7-n+2n\beta}{7-n-2\beta}.
\end{equation}
The instability condition, therefore, becomes
\begin{equation}
\frac{(7-n)[(2\beta -9)n-1]}{8n(n+2\beta -7)}<0.
\end{equation}
We note that since the stress exponent is within the range $0<n\leq 1$, the above instability condition is simplified to
\begin{equation}
\frac{(2\beta -9)n-1}{n+2\beta -7}<0.
\end{equation}
For $\beta=2$, however, this instability condition is violated because its  left hand side becomes $(1+5n)/(3-n)$ which is always positive for the allowed range of the stress exponent.  It means that our solutions are dynamically stable subject to the thermal-viscous instability. 

\section{Importance of the Radiative Heating}
We have so far neglected radiative heating sources (e.g., central star irradiation, cosmic rays) in our analysis. However, these extra heating sources are important in the thermodynamics of the PPDs. In this work, we consider irradiation by a central star.  If we denote radiative heating by $\Gamma_{\rm rad}$, this new term is added to the right-hand side of the energy equation (\ref{eq:m2}). Radiative heating is a complicated function of the disc quantities, however, we can use the following relation \citep[e.g.,][]{Armi,frank}
\begin{equation}\label{eq:rad-heating}
\Gamma_{\rm rad} = \Gamma_{\rm rad,0} (\frac{r}{r_0})^{-2},
\end{equation}
where  $\Gamma_{\rm rad,0}=2g(H/r)  L_{\star}/4\pi r_{0}^2$ and $g=(d{\rm ln} H/d{\rm ln}r) -1 $ and $L_{\star}$ is the star luminosity. The coefficient $g$ lies between $1/8$ and $2/7$ and the ratio $H/R$ is approximately constant in a disc \citep[see p. 130 in][]{frank}.   

\begin{figure}
\includegraphics[scale=0.6]{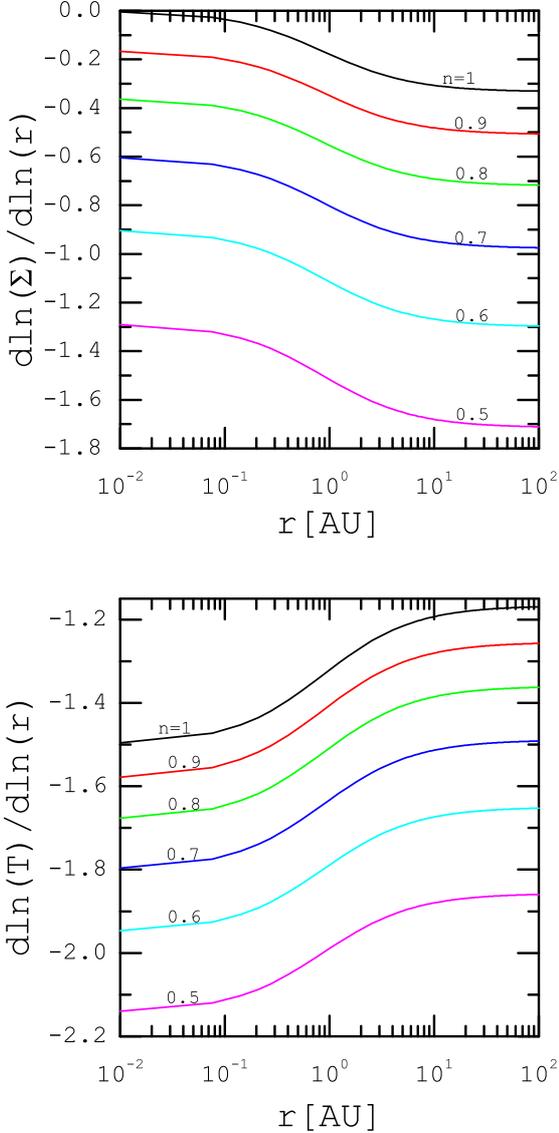}
\caption{Radial slope of the surface density (top) and temperature (bottom) as a function of the radius in the optically thick  regime for different values of the stress exponent, as labeled. Here, we have $M_{\star}=M_{\odot}$, $\dot{M}=10^{-8} {\rm M}_{\odot}/{\rm yr}$, $\alpha=0.01$, $\Sigma_0 = 10$ g cm$^{-2}$ and $T_0 =10$ K, and, the opacity exponent is $\beta=2$. Luminosity of the central star is $L_{\star}=3.9\times 10^{33}$ erg s$^{-1}$ and we assume that $g=1/8$ and $H/r\simeq 0.01$.  Thus, we obatin  $\zeta =1.16$.}\label{fig:f4}
\end{figure}

We can explore the properties of the optically thick solutions in the presence of radiative heating. To this end, we add equation (\ref{eq:rad-heating}) to the right-hand side of the energy equation (\ref{eq:m2}). Then, the obtained modified energy equation and equations (\ref{eq:m1}), (\ref{eq:m3}), and (\ref{eq:cs}) would enable us to obtain disc quantities:
\begin{displaymath}
\frac{\Sigma}{\Sigma_0} = {\cal A} \alpha^{\delta_1} \left (\frac{\dot{M}}{\dot{M}_0} \right )^{\delta_2} \left (\frac{M_\star}{M_{\odot}} \right )^{\delta_3} \left (\frac{r}{r_0} \right )^{\delta_4} 
\end{displaymath}
\begin{equation}\label{eq:densH-a}
\times \left [ 1+\zeta \left (\frac{r}{r_0} \right ) \right ]^{-\frac{n+1}{1+(9-2\beta )n}},
\end{equation}
\begin{displaymath}
\frac{T}{T_0}= {\cal B} \alpha^{\delta_5} \left (\frac{\dot{M}}{\dot{M}_0} \right )^{\delta_6 }  \left (\frac{M_\star}{M_{\odot}} \right )^{\delta_7} \left (\frac{r}{r_0} \right )^{\delta_8}  
\end{displaymath}
\begin{equation}\label{eq:tempH-a}
\times \left [ 1+\zeta \left (\frac{r}{r_0} \right ) \right ]^{\frac{2n}{1+(9-2\beta )n}},
\end{equation}
where dimensionless parameter $\zeta$ denotes the ratio of the radiative heating and viscous heating coefficients, i.e. 
\begin{equation}
\zeta = \frac{\Gamma_{\rm rad, 0}}{\frac{3}{8\pi} \dot{M}_0 \Omega_{0}^2} (\frac{\dot{M}}{\dot{M}_0})^{-1}(\frac{M_\star}{M_{\odot}})^{-1}.
\end{equation}
Obviously, if we set $\zeta=0$, the above solutions reduce to the optically thick solutions without radiative heating, i.e. equations (\ref{eq:dens-a}) and (\ref{eq:temp-a}).

In Figure \ref{fig:f4} we explore radial slope of the surface density (top) and temperature (bottom) as a function of the radius for different values of the stress exponent in the presence of radiative heating. We consider a solar mass star with a luminosity $L_{\star}=L_{\odot}=3.9\times 10^{33}$ erg s$^{-1}$. The other model parameters are $M_{\star}=M_{\odot}$, $\dot{M}=10^{-8} {\rm M}_{\odot}/{\rm yr}$, $\alpha=0.01$, $\Sigma_0 = 10$ g cm$^{-2}$ and $T_0 =10$ K, and, the opacity exponent is $\beta=2$. We then obtain $\zeta=1.16$. The first point to note is that, regardless of the stress exponent, the distribution of surface density in the outer region is steeper compared to the inner region. The difference in the slope of the surface density of the inner and outer regions increases with decreasing the stress exponent. Regarding the temperature slope, the situation is the opposite, i.e. as we go to the outer parts of a disc, the slope decreases.

\section{Time-dependent Model}

In the previous sections, we explored the structure of a steady-state thin accretion disc with a power-law stress-pressure relationship. An accretion disc, however, is not necessarily in a steady-state and its evolution can be studied using analytical or numerical solutions \citep[e.g.,][]{lin82,ruden86,lin87,cannizzo,hartmann98,step98,step98b,rice2009}. Most of the previous studies on the time-evolution of an accretion disc adopted the standard $\alpha -$formalism for the viscosity. They found that the accretion rate quickly becomes independent of the radial distance \cite[e.g.,][]{ruden86,step98,rice2009}. Using this typical behavior of the accretion rate, as an approximation, \cite{chambers} obtained interesting time-dependent analytical solutions for the structure of an accretion disc. The advantage of this approach which has already been developed by \cite{step98, step98b} is its simplicity and the time-dependent behavior of the disc quantities can be explored analytically. 

Here, we apply a similar method, but for a disc with a power-law stress-pressure relationship. When a change in the mass accretion rate occurs on a longer timescale than the disc viscous timescale, we can stitch together our steady-state solutions with a time-dependent accretion rate and the constraints due to the total disk mass and angular momentum conservation. This approximation is violated when viscous timescale becomes comparable to the timescale of change in the accretion rate. We note that the viscous timescale increases outwards through the disc. Our implemented approximation, thereby, is valid in the inner disc and it is not going to hold in the outer region of a disc.

We have to determine the total mass of the disc, $M_{\rm d}$, and the total angular momentum of the disc, $L$. The accretion rate is assumed to be only a function of time, however, equation (\ref{eq:m1}) is still applicable as the disc evolves. Therefore, the surface density and the temperature solutions (\ref{eq:dens-a}) and (\ref{eq:temp-a}) are valid in the time-dependent case, but the accretion rate is no longer independent of time. In other words, time-dependence of the solutions comes from the accretion rate which is a function of time. By assuming that reduction of the disc total mass is solely due to the accretion onto the central object along with the conservation of the total angular momentum will eventually lead to a differential equation for the accretion rate \citep{chambers}. Having the accretion rate as a function of time, we can  determine the temporal dependence of the disc quantities. 

The total mass of the disc is written as
\begin{equation}
M_{\rm d} = 2\pi s_{0}^2 \Sigma_0 \int_{\frac{s_{\rm in}}{s_0 }}^{\frac{s}{s_0 }} (r/s_0 ) (\Sigma /\Sigma_0 ) d(r/s_0 ) ,
\end{equation}
where $s_0$ is the initial size of the disc. Furthermore, the outer radius of the disc at time $t$ is $s$ and the inner edge of the disc is denoted by $s_{\rm in}$. Using equation (\ref{eq:dens-a}) and assuming $s_{\rm in}=0$, for simplicity, the total mass becomes
\begin{equation}\label{eq:Md}
M_{\rm d} = 2\pi s_{0}^2 \Sigma_{0} {\cal A}_{\rm thick} \alpha^{\delta_1} \dot{m}^{\delta_2} M_{1}^{\delta_3} (1+ \delta_4 )^{-1} (\frac{s}{s_0})^{2+ \delta_4}.
\end{equation}
Moreover, the total angular momentum is written as
\begin{equation}
L=2\pi \sqrt{G M_\star} s_{0}^{5/2} \Sigma_0 \int_{\frac{s_{\rm in}}{{s_0}}}^{\frac{s}{s_0}} (r/s_0 )^{3/2} (\Sigma / \Sigma_0 ) d(r/s_0),
\end{equation}
and upon substituting from equation (\ref{eq:dens-a}), we then obtain
\begin{equation}\label{eq:L}
L = 2\pi \sqrt{G M_\star} s_{0}^{5/2} \Sigma_{0} {\cal A}_{\rm thick} (\delta_4 + \frac{5}{2})^{-1} \alpha^{\delta_1} \dot{m}^{\delta_2} M_{1}^{\delta_3 } (\frac{s}{s_0 })^{\frac{5}{2}+\delta_4}.
\end{equation}

\begin{figure}
\includegraphics[scale=0.6]{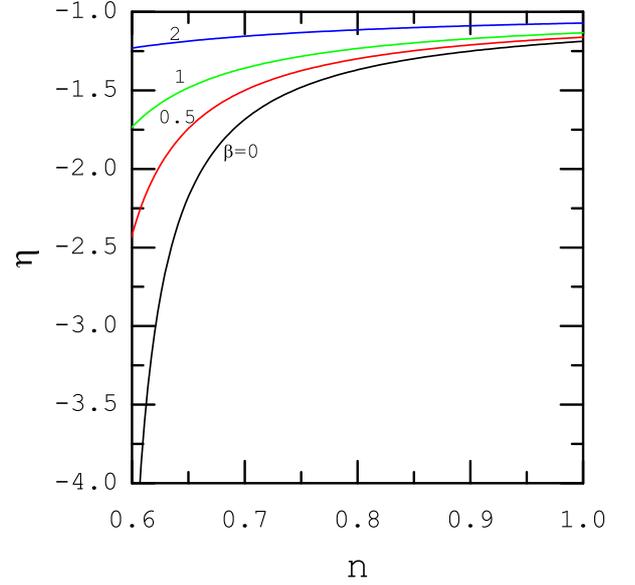}
\caption{Profile of the accretion exponent $\eta$ as a function of the exponent $n$. Each curve is labeled with the corresponding opacity exponent $\beta$.}\label{fig:f5}
\end{figure}

\begin{figure}
\includegraphics[scale=0.6]{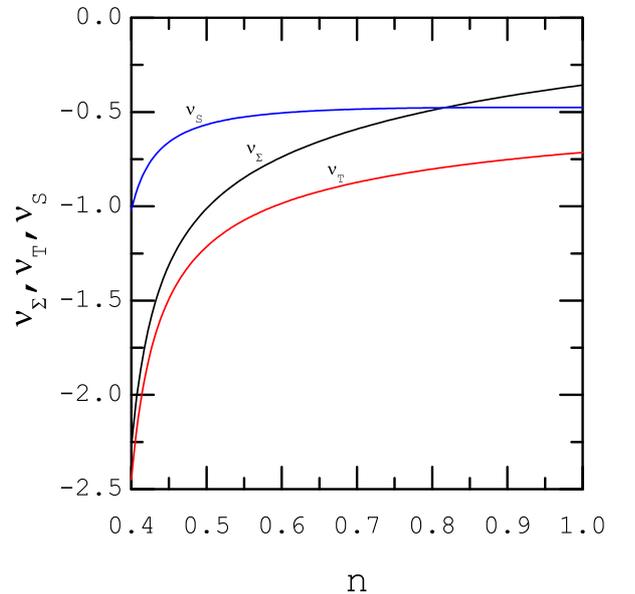}
\caption{Profiles  of the exponents $\nu_{\Sigma}$,  $\nu_{\rm T}$ and $\nu_{\rm S}$ as a function of the exponent $n$. The opacity exponent is $\beta =2$.}\label{fig:f6}
\end{figure}

Note that in the above equation the accretion rate is a function of time, thought its time-dependence is still unknown. By eliminating the accretion rate $\dot{m}$ between equations (\ref{eq:Md}) and (\ref{eq:L}), we therefore obtain the following relation, 
\begin{equation}\label{eq:MdL}
\frac{M_{\rm d}}{L} = \frac{1}{\sqrt{G M_{\star } s_0 }} \frac{\frac{5}{2} + \delta_4 }{2+\delta_4} (\frac{s}{s_0})^{-1/2},
\end{equation}
where $L$ is a conserved quantity. Obviously, we have $s(t=0)=s_0$ and the initial mass of the disc, $M_{\rm 0d}$, becomes
\begin{equation}
M_{\rm 0d} = \frac{L}{\sqrt{GM_\star } s_0} \frac{\frac{5}{2} +\delta_4 }{2+\delta_4 }.
\end{equation}
Using the above equation, we can then simplify equation (\ref{eq:MdL}) as follows
\begin{equation}\label{eq:s-Md}
\frac{s}{s_0} = (\frac{M_{\rm d}}{M_{\rm 0d}})^{-2}.
\end{equation}

Since the total angular momentum is conserved, from equation (\ref{eq:L}) we obtain
\begin{equation}\label{eq:L-con}
\frac{\dot{M}}{\dot{M}_0 } = (\frac{s}{s_0})^{-\frac{\frac{5}{2} +\delta_4 }{\delta_2 }}.
\end{equation}
Using equations (\ref{eq:s-Md}) and (\ref{eq:L-con}), a power-law relation between the accretion rate and the total mass of the disc is obtained, i.e.
\begin{equation}\label{eq:Mdot-Md}
\frac{\dot{M}}{\dot{M}_0} = (\frac{M_{\rm d}}{M_{\rm 0d}})^{\frac{5+2\delta_4}{\delta_2}}.
\end{equation}

We assume that the total mass of disc decreases only due to the mass accretion onto the central object, i.e.
\begin{equation}
\frac{dM_{\rm d}}{dt} = -\dot{M},
\end{equation}
or
\begin{equation}
\frac{dM_{\rm d}}{dt} = - \dot{M}_0 (\frac{M_{\rm d}}{M_{\rm 0d}})^{\frac{5+2\delta_4}{\delta_2}} .
\end{equation}
This is the main differential equation which gives the total mass as a function of time, i.e.
\begin{equation}\label{eq:Md-t}
\frac{M_{\rm d}}{M_{\rm 0d}} = (1+\frac{t}{t_0})^{-\frac{\delta_2}{2\delta_4 - \delta_2 +5}},
\end{equation}
where
\begin{equation}
t_{0} = \delta_2 (5+2\delta_4 - \delta_2 )^{-1} ( M_{\rm 0d} / \dot{M}_0 ).
\end{equation}
Upon substituting equation (\ref{eq:Md-t}) into equation (\ref{eq:Mdot-Md}), we thereby arrive at the following relation for the accretion rate as a function of time:
\begin{equation}\label{eq:Mdot-Time}
\frac{\dot{M}}{\dot{M}_0} = (1+\frac{t}{t_0})^{\eta},
\end{equation}
where the exponent $\eta$ is
\begin{equation}
\eta = -\frac{5+2\delta_4}{5+2\delta_4 - \delta_2}=-\frac{16n\beta-75n-12\beta+37}{2(8n\beta-38n-7\beta+22)}.
\end{equation}

Figure \ref{fig:f5} displays the exponent of the accretion rate $\eta$ as a function of the stress exponent $n$ for different values of $\beta$. It shows that the accretion rate gradually decreases with the age of the disc, however, its decay rate strongly depends in a non-trivial way on the exponent $n$. If the exponent $n$ is adopted less than around 0.5, however, we showed that the slopes of the surface density and temperature tend to unphysically large values which are irrelevant for describing the structure of a PPD. Therefore, the analysis in the time-dependent case is restricted to the values of $n$ greater than 0.5. We found that the variation of the accretion rate with time is faster as the value of the exponent $n$ decreases. However, the dependence of $\eta$ on the exponent $n$ becomes weaker as the opacity exponent increases. Many authors have already derived the decay of the accretion rate with time as a power-law function using numerical or analytical solutions under certain simplifying assumptions \citep[e.g.,][]{Lynden,lin82,filipov,cannizzo,king98,lipu2000,tanaka2011,Lipu15}. The obtained exponent $\eta$ based on their models, however, strongly depends on the adopted viscosity, opacity and even the imposed boundary conditions. Observational evidence, on the other hand, has also confirmed a power-law decay for the accretion rate in some of the accreting systems. For instance, \cite{hartmann98} obtained exponent $\eta$ between -1.5 and -2.8 by analyzing T-Tauri stars. They prescribed turbulent viscosity simply as a power-law function of the radial distance. If we adopt this range of variations for the accretion decay exponent, our analysis shows that the stress exponent $n$ is between 0.74 and 0.62 when the disc is optically thick and the opacity exponent is $\beta =2$. If we adopt a larger value for the opacity exponent, the allowed range of the stress exponent is slightly modified. For instance, we have $0.54\leq n \leq0.64$, when $\beta =1$. If we set $n=1$, the obtained accretion slope $\eta$ is not consistent with the  findings of \cite{hartmann98}.

Now, we can examine how the density and temperature distributions are changing over the time. As we mentioned earlier, the temporal dependence of the disc quantities is through the accretion rate. Upon substituting equation (\ref{eq:Mdot-Time}) into equations (\ref{eq:dens-a}) and (\ref{eq:temp-a}), therefore, we obtain $\Sigma \propto (1+t/t_0 )^{\nu_{\Sigma}}$ and $T \propto (1+t/t_0 )^{\nu_{\rm T}}$, where $\nu_{\Sigma}=\eta \delta_2$ and $\nu_{\rm T}=\eta \delta_6$. Figure \ref{fig:f6} displays variations of these exponents as a function of $n$ when the opacity exponent is $\beta =2$. As the stress exponent $n$ decreases, the reduction of the surface density and temperature with the age of disc become faster. For a given parameter $n$,  however, the temperature declines much faster than the surface density. Although for a case with $n=1$ the difference between $\nu_{\Sigma}$ and $\nu_{\rm T}$ is not very significant, the difference between $\nu_{\Sigma}$ and $\nu_{\rm T}$ becomes larger if the stress exponent decreases.

By substituting equation (\ref{eq:Mdot-Time}) into equation (\ref{eq:snow}), we obtain $r_{\rm snow} \propto (1+t/t_0 )^{\nu_{\rm S}} $, where $\nu_{\rm S}=-\eta \delta_{6}/\delta_8$. Figure \ref{fig:f5} exhibits the profile of the exponent $\nu_{\rm S}$ as a function of the exponent $n$ for a case with the opacity exponent $\beta =2$. As long as the stress exponent $n$ is between around 0.75 and 1, the migration rate of the snow-line is more or less independent of the exponent $n$. This figure shows that for $0.75\leq n \leq 1$, we have $\nu_{\rm S} \simeq -0.6$ to $-0.5$. However, once the exponent $n$ becomes smaller than around 0.75, migration rate of the snow-line becomes very fast so that for $n\simeq 0.6$, the exponent $\nu_{\rm S}$ reaches to -2.

\section{Discussions}
Following numerical simulations of the magnetized accretion discs \citep{Sano,Minoshima,Ross} which showed that the radial-azimuthal component of the stress tensor due to MRI is a power-law function of the gas pressure, we constructed steady-state and time-dependent accretion disc models to explore the physical consequences of this new prescribed stress-pressure relation within the framework of the standard disc model. We found that the structure of a disc with a power-law stress-pressure relationship strongly depends on the stress exponent, however, as this exponent becomes less than around 0.5, the radial profiles of the disc quantities become so steep that this trend is very unlikely to be confirmed by the observations. It means that one can constrain the stress exponent $n$ using our analytical solutions. 

We also considered radiative heating and presented analytical solutions for a PPD when both radiative heating and viscous heating are considered. Although the radiative heating is treated in a parameterized form, the most prominent influence of the radiative heating on the radial profile of the disc quantities is shown to occur for smaller values of the stress exponent.  

As \cite{Ross} demonstrated based on their simulations, the exponent $n$ depends on the imposed  net vertical flux of the magnetic field. However, they did not provide a possible approximate relation between the stress exponent and the imposed magnetic field which could  be quantified in terms of the ratio of the gas pressure and magnetic pressure. Using such a relation (if any), it is possible to extend our analysis to the magnetized discs.

 As we mentioned earlier, the opacity is approximated as a power-law function of the temperature and its exponent, $\beta$, depends on the temperature range. Although we presented the solutions in their general forms, most of the discussions were based on the solutions associated with a particular case, i.e. $\beta =0$. This simplification enabled us to focus on the influence of the stress exponent. A realistic model is to adopt solutions depending on the temperature range, and to this extent, an iterative procedure is needed because we do not initially know the temperature and it is something which the model aims at deriving. A similar approach has already been used by other authors \citep[e.g.][]{fendt}, but in their model  relied on a linear stress-press relationship. Using our analytical solutions with a power-law stress-pressure relationship, one can construct piecewise solutions corresponding to the considered temperature ranges.

Although we applied our analytical solutions to the PPDs, there are considerable uncertainties about the true nature of the angular momentum transport at radii beyond the inner region of a PPD. While the MRI will probably work at the radii less than a few AUs, outside of that it is uncertain what is happening. Even if MRI is still effective at these regions, it is very unlikely a single exponent $n$ is adequate for describing the entire structure of a PPD. Considering these complexities in mind, we think, the derived solutions are applicable to the inner parts of a PPD so long as MRI is the dominant driver of angular momentum transport.

In summary, since the classical $\alpha-$formalism serves as a foundation for many theoretical studies of accretion discs, it would be important to examine the consequences of a power-law stress-pressure relationship in other accreting systems.


\section*{Acknowledgements}
We are very grateful to the referee for a very constructive and thoughtful report that greatly helped us to improve the paper. MS is grateful to Rebecca G. Martin and Henrik Latter for their constructive comments on an early version of this paper. This work has been supported financially by Research Institute for Astronomy \& Astrophysics of  Maragha (RIAAM) under research project No. 1/5440-12.

\bibliographystyle{mnras}
\bibliography{reference} 




\bsp	
\label{lastpage}
\end{document}